\documentclass[]{spie}  

\usepackage{amsmath,amsfonts,amssymb}
\usepackage{graphicx}
\usepackage[colorlinks=true, allcolors=blue]{hyperref}

\title{Design and characterization of the SPT-3G receiver}

\author[a,b]{J. A. Sobrin}
\author[c]{P. A. R. Ade}
\author[d,e]{Z. Ahmed}
\author[f,a]{A. J. Anderson}
\author[g]{J. S. Avva}
\author[a]{R. Basu Thakur}
\author[h,a]{A. N. Bender}
\author[f,a,i]{B. A. Benson}
\author[a,j,b,h,i]{J. E. Carlstrom}
\author[h,a]{F. W. Carter}
\author[h]{T. W. Cecil}
\author[h,a,i]{C. L. Chang}
\author[k]{J. F. Cliche}
\author[g]{A. Cukierman}
\author[g]{T. de Haan}
\author[l]{J. Ding}
\author[k,m]{M. A. Dobbs}
\author[a,b]{D. Dutcher}
\author[n]{W. Everett}
\author[o]{A. Foster}
\author[a,p]{J. Gallicchio}
\author[k]{A. Gilbert}
\author[g]{J. C. Groh}
\author[g]{S. T. Guns}
\author[n,q]{N. W. Halverson}
\author[h,r]{A. H. Harke-Hosemann}
\author[g]{N. L. Harrington}
\author[h,a]{J. W. Henning}
\author[g]{W. L. Holzapfel}
\author[g]{N. Huang}
\author[d,s,e]{K. D. Irwin}
\author[g]{O. B. Jeong}
\author[f]{M. Jonas}
\author[l]{T. S. Khaire}
\author[r]{A. M. Kofman}
\author[o]{M. Korman}
\author[f]{D. L. Kubik}
\author[h]{S. Kuhlmann}
\author[d,s,e]{C.-L. Kuo}
\author[g,t]{A. T. Lee}
\author[a]{A. E. Lowitz}
\author[a,j,b,i]{S. S. Meyer}
\author[u]{D. Michalik}
\author[k]{J. Montgomery}
\author[r]{A. Nadolski}
\author[v]{T. Natoli}
\author[f]{H. Nguyen}
\author[k]{G. I. Noble}
\author[l]{V. Novosad}
\author[a]{S. Padin}
\author[a,b]{Z. Pan}
\author[l]{J. Pearson}
\author[l]{C. M. Posada}
\author[a,b]{W. Quan}
\author[f,a]{A. Rahlin}
\author[o]{J. E. Ruhl}
\author[n]{J.T. Sayre}
\author[a,i]{E. Shirokoff}
\author[w]{G. Smecher}
\author[x]{A. A. Stark}
\author[d,s]{K. T. Story}
\author[t]{A. Suzuki}
\author[d,s,e]{K. L. Thompson}
\author[c]{C. Tucker}
\author[v,y]{K. Vanderlinde}
\author[r,z]{J. D. Vieira}
\author[h]{G. Wang}
\author[aa,g]{N. Whitehorn}
\author[h]{V. Yefremenko}
\author[d,s,e]{K. W. Yoon}
\author[y]{M. R. Young}

\affil[a]{Kavli Institute for Cosmological Physics, University of Chicago, 5640 South Ellis Avenue, Chicago, IL, USA 60637}
\affil[b]{Department of Physics, University of Chicago, 5640 South Ellis Avenue, Chicago, IL, USA 60637}
\affil[c]{School of Physics and Astronomy, Cardiff University, Cardiff CF24 3YB, United Kingdom}
\affil[d]{Kavli Institute for Particle Astrophysics and Cosmology, Stanford University, 452 Lomita Mall, Stanford, CA, USA 94305}
\affil[e]{SLAC National Accelerator Laboratory, 2575 Sand Hill Road, Menlo Park, CA, USA 94025}
\affil[f]{Fermi National Accelerator Laboratory, MS209, P.O. Box 500, Batavia, IL, USA 60510}
\affil[g]{Department of Physics, University of California, Berkeley, CA, USA 94720}
\affil[h]{High-Energy Physics Division, Argonne National Laboratory, 9700 South Cass Avenue., Argonne, IL, USA 60439}
\affil[i]{Department of Astronomy and Astrophysics, University of Chicago, 5640 South Ellis Avenue, Chicago, IL, USA 60637}
\affil[j]{Enrico Fermi Institute, University of Chicago, 5640 South Ellis Avenue, Chicago, IL, USA 60637}
\affil[k]{Department of Physics and McGill Space Institute, McGill University, 3600 Rue University, Montreal, Quebec H3A 2T8, Canada}
\affil[l]{Materials Sciences Division, Argonne National Laboratory, 9700 South Cass Avenue, Argonne, IL, USA 60439}
\affil[m]{Canadian Institute for Advanced Research, CIFAR Program in Cosmology and Gravity, Toronto, ON, M5G 1Z8, Canada}
\affil[n]{CASA, Department of Astrophysical and Planetary Sciences, University of Colorado, Boulder, CO, USA 80309}
\affil[o]{Department of Physics, Center for Education and Research in Cosmology and Astrophysics, Case Western Reserve University, Cleveland, OH, USA 44106}
\affil[p]{Harvey Mudd College, 301 Platt Boulevard., Claremont, CA, USA 91711}
\affil[q]{Department of Physics, University of Colorado, Boulder, CO, USA 80309}
\affil[r]{Department of Astronomy, University of Illinois at Urbana-Champaign, 1002 West Green Street, Urbana, IL, USA 61801}
\affil[s]{Deptartment of Physics, Stanford University, 382 Via Pueblo Mall, Stanford, CA, USA 94305}
\affil[t]{Physics Division, Lawrence Berkeley National Laboratory, Berkeley, CA, USA 94720}
\affil[u]{University of Chicago, 5640 South Ellis Avenue, Chicago, IL, USA 60637}
\affil[v]{Dunlap Institute for Astronomy \& Astrophysics, University of Toronto, 50 St. George Street, Toronto, ON, M5S 3H4, Canada}
\affil[w]{Three-Speed Logic, Inc., Vancouver, B.C., V6A 2J8, Canada}
\affil[x]{Harvard-Smithsonian Center for Astrophysics, 60 Garden Street, Cambridge, MA, USA 02138}
\affil[y]{Department of Astronomy \& Astrophysics, University of Toronto, 50 St. George Street, Toronto, ON, M5S 3H4, Canada}
\affil[z]{Department of Physics, University of Illinois Urbana-Champaign, 1110 West Green Street, Urbana, IL, USA 61801}
\affil[aa]{Department of Physics and Astronomy, University of California, Los Angeles, CA, USA 90095}

\authorinfo{Corresponding Author: J. A. Sobrin (joshuasobrin@uchicago.edu)}

\begin{document} 
\maketitle

\begin{abstract}
The SPT-3G receiver was commissioned in early 2017 on the 10-meter South Pole Telescope (SPT) to map anisotropies in the cosmic microwave background (CMB).  New optics, detector, and readout technologies have yielded a multichroic, high-resolution, low-noise camera with impressive throughput and sensitivity, offering the potential to improve our understanding of inflationary physics, astroparticle physics, and growth of structure.  We highlight several key features and design principles of the new receiver, and summarize its performance to date.
\end{abstract}


\section{INTRODUCTION}
\label{sec:intro}  

The 10-meter South Pole Telescope (SPT) surveys the cosmic microwave background (CMB) with arcminute angular resolution from the Amundsen-Scott South Pole Station\cite{Carlstrom11}.  The South Pole provides a transparent, stable atmosphere with observation fields that never set.  Additionally, the large primary mirror of the SPT provides high-angular resolution imaging capability, allowing receivers on the SPT to produce sensitive maps of temperature- and polarization-anisotropies of the CMB across a broad range of angular scales.  These deep and high-resolution maps can be used to better constrain CMB power spectra extending to high multipoles.  More precise measurements of the CMB will improve our understanding of cosmological models and parameters, including the energy-scale of inflation, the sum of neutrino masses, and the dark energy equation of state.

Over the past several years, CMB instruments have achieved better sensitivity primarily through increasing the number of photon-noise-limited detectors populating their focal planes.  SPT-3G is the third-generation CMB experiment commissioned on the SPT, and achieved first light in early 2017.  The instrument features an array of $\sim$16,000 temperature- and polarization-sensitive detectors---an order of magnitude increase over its predecessor, SPTpol.  The dramatic increase in detector-count was made possible by (i) a restructured optics design yielding a larger field-of-view (FOV) with 3-band (95, 150, 220 GHz) frequency coverage, (ii) the development of tri-chroic dual-polarization pixels, and (iii) an improved frequency-domain SQUID readout system capable of 68x multiplexing.  An overview of the experiment and its scientific goals can be found in Ref.~\citenum{Benson14}.

In this manuscript we describe features of the cryogenic receiver which successfully facilitated these design goals.  We also present measurements of optical and cryogenic performance to date for the new receiver.  Further details on integrated performance over the first two years and recent science projections can be found in Refs.~\citenum{Anderson18} and~\citenum{Bender18}.

\section{TELESCOPE DESIGN}
\label{sec:telescope}
   \begin{figure} [ht]
   \begin{center}
   \begin{tabular}{c}
   \includegraphics[height=7.0cm]{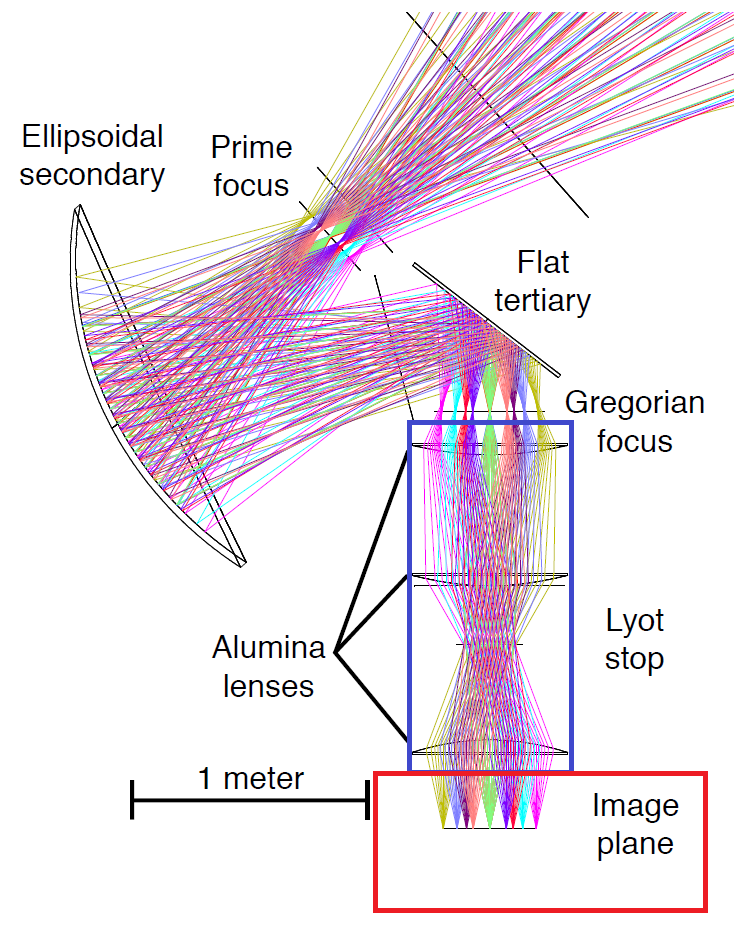}
   \includegraphics[height=6.75cm]{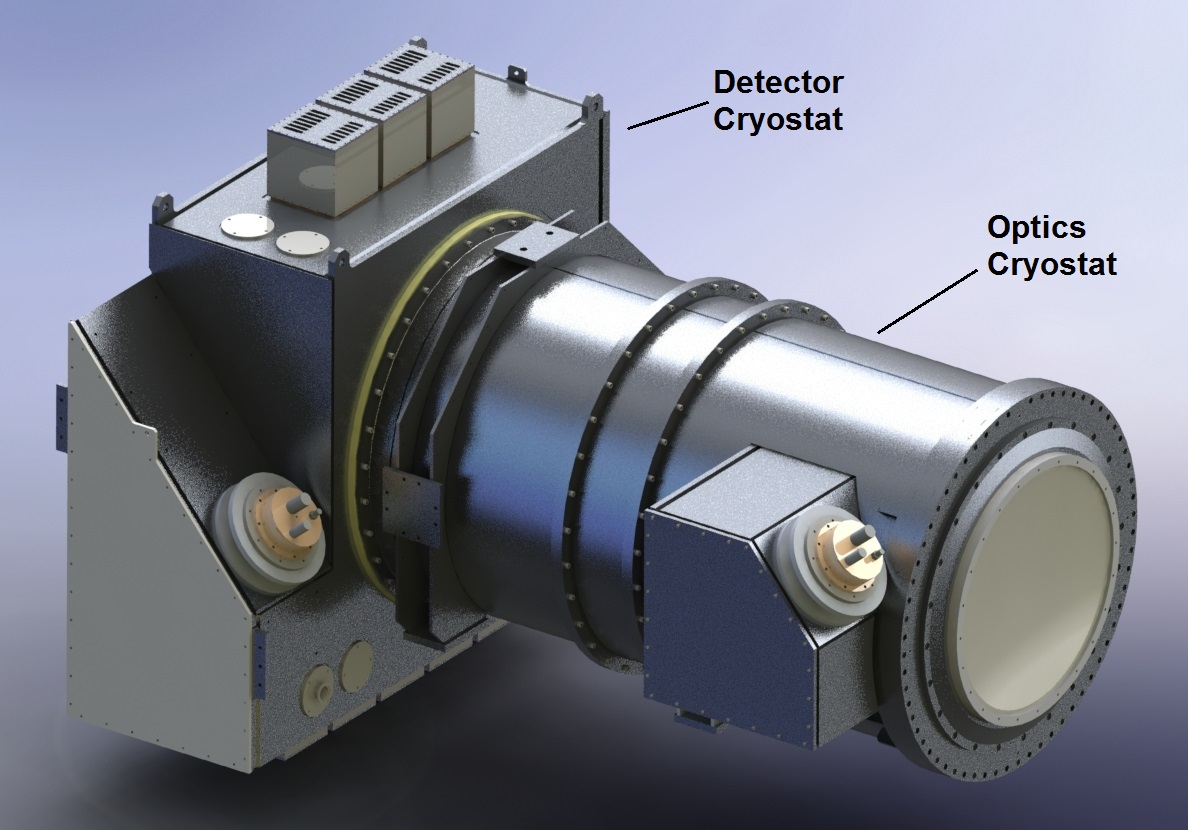}
   \end{tabular}
   \end{center}
   \caption[telescope] 
   { \label{fig:telescope} 
(Left) Ray-trace diagram of the SPT-3G optics design, excluding the primary mirror.  The blue and red borders denote the optics and detector cryostats, respectively.  (Right) A CAD rendering of the SPT-3G receiver with the optics and detector cryostats coupled together for integrated operation.}
   \end{figure}

SPT-3G has modified the optics design to achieve a significantly wider FOV than the previous SPT-SZ and SPTpol cameras, while retaining the off-axis Gregorian design of the SPT\cite{Padin08}.  A new 2-meter ellipsoidal secondary and 1-meter flat tertiary mirror fabricated by Cospal Antennas\footnote{http://www.cospal.it/} are used to form the Gregorian focus immediately outside the receiver, as shown in Fig. 1.  A set of refractive optics, held at 4 K to reduce thermal emission, re-images the Gregorian focus to a flat image plane at which the cryogenic detector array is positioned.  This design provides a 1.9 degree diameter FOV across a 430 mm focal plane, a significant increase over the $\sim$1 degree diameter FOV provided by the SPT-SZ and SPTpol optics designs.  On top of superior throughput, the design achieves excellent image quality, with predicted Strehl ratios greater than 0.98 across the entire detector array in all three frequency bands, with a full width at half maximum (FWHM) of 1.6, 1.2, and 1.1 arcminutes at 95, 150, and 220 GHz, respectively.

SPT-3G uses high-purity alumina lenses that are manufactured by Coorstek, Inc.\footnote{https://www.coorstek.com/}  The lenses are sintered using 99.5\% pure alumina powder, and weigh roughly 100 pounds each.  Alumina was chosen as a material because it can be manufactured at the large diameter (720 mm) required by the SPT-3G optics design, is relatively high-index and low-loss at mm-wavelengths, and maintains better thermal conductivity at cryogenic temperatures than other candidate plastic materials.  Using a Fourier transform spectrometer (FTS), we have measured witness samples of the lenses to have a refractive index of 3.10 and loss tangent of $1.6 \times 10^{-4}$.

Figure 1 includes an image of the SPT-3G receiver, which has an optics (blue outline) and detectors (red outline) section.  The design allows for both portions to be cooled and tested independently as standalone cryostats for characterization, but in their observing configuration, both cryostats share the same vacuum space as shown in Fig 2.  The nominal 50 K and 4 K stages are cooled using a pair of Cryomech\footnote{http://www.cryomech.com/} PT-415 refrigerators, with the cold stages and associated components mechanically supported and thermally isolated via a set of G10 struts.  The cryostats were largely fabricated by Dial Machine, Inc.\footnote{https://www.dialmachine.com/}  The integrated cryostat exhibited a leak rate of $4 \times 10^{-5}$ mbar-liters/second (estimated from the base pressure of the cryostat at room-temperature after the first year of observations) which is consistent with the leak rate expected from the O-rings.

\section{OPTICS CRYOSTAT}

The optics cryostat primarily positions and cools the infrared (IR) blocking filters, lenses, and Lyot stop of the SPT-3G optics design.  Past the ambient-temperature reflective optics and Gregorian focus, light enters the cryogenic receiver through a 30-mm-thick HDPE vacuum window with a diameter of 27 inches.  Suspended from this window assembly is a series of 0.125-inch-thick Zotefoam sheets which absorptively filter thermal IR radiation.  The sheets are thermally insulated from one another using G10 spacers, causing each filter to sit slightly colder than the prior due to radiative heat-transfer (similar to the principle behind multi-layer insulation).  At the nominal 50 K stage, a 0.6-inch-thick flat disk of alumina acts as a final cold absorptive low-pass filter before the 4 K lenses.  This flat alumina plate is mechanically supported and cooled along the perimeter by a copper ring, which is suspended from the vacuum window assembly by thermally-insulating G10 struts, and is directly connected to the first-stage cold-head of the optics cryostat pulse-tube cooler (PTC).  Thermometers located on this alumina filter measure an average operating temperature of 50 K, with a differential of less than 5 K across the plate during observations.

The ultimate sensitivity of a CMB instrument is significantly affected by the power the instrument itself deposits on the detectors through thermal emission, as well as the overall optical efficiency of the integrated system.  The ability to successfully cool optics elements and minimize reflections through the cryostat was crucial for achieving a well-understood low-noise instrument.  Below we highlight several techniques which helped to realize these goals.

\vspace{1cm}

   \begin{figure} [ht]
   \begin{center}
   \begin{tabular}{c}
   \includegraphics[height=12cm]{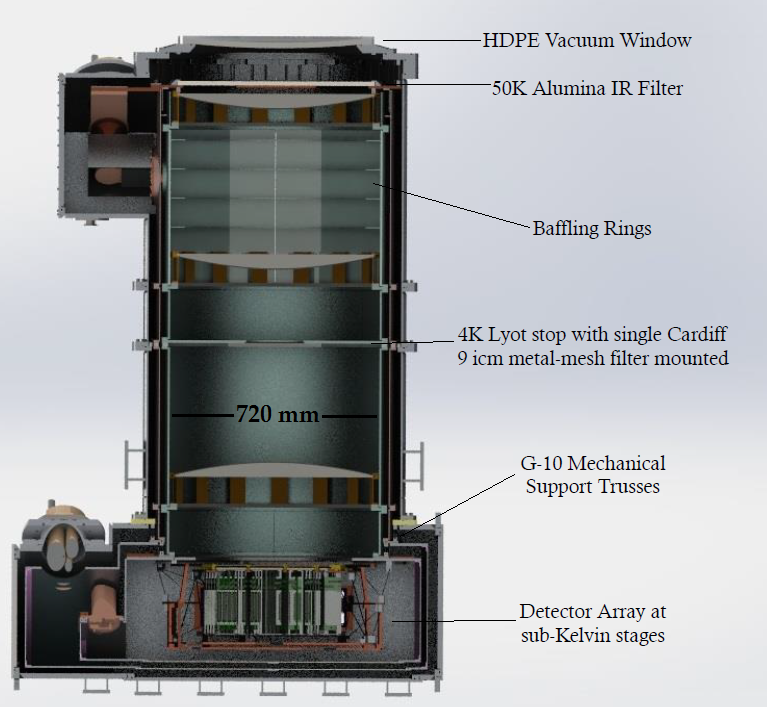}
   \end{tabular}
   \end{center}
   \caption[optics_cutaway] 
   { \label{fig:optics_cutaway} 
Cross-sectional view of the SPT-3G receiver architecture.}
   \end{figure}

\subsection{Controlling Stray Reflections}

An ideal optics system should aim to mitigate the optical coupling of detectors to hot and emissive surfaces, as well as achieve a predictable point spread function.  Suppressing undesirable reflections and scattering within the receiver facilitates these two goals.

The Lyot stop provides a cold surface which terminates power from the outer parts of the beam.  Near the window of the receiver, baffling rings mitigate grazing reflections that might terminate on hot surfaces outside the cryogenic volume, yet not on the sky.  Most importantly, all of the potentially reflective metallic surfaces inside the 4 K volume are completely coated with some form of cold microwave absorbing material, most often Eccosorb HR-10 when possible.  These absorbers were epoxied to the surfaces using a thin layer of Stycast 2850FT.

\subsection{Minimizing Thermal Gradients}

Many important cryogenic elements in the optics cryostat are not located physically near the PTC, often separated by a long series of thermal resistances, and significant thermal gradients could develop across components.  A failure to minimize these gradients would prevent optics components from achieving their optimally cold base temperatures, resulting in elevated emissive power.  Efforts were taken to minimize the loading from the optics elements by improving the thermal conductivity between them and the PTC.

\subsubsection{High-purity aluminum}

Aluminum 1100 is used for the 50 K and 4 K radiation shields due to its superior thermal conductivity over aluminum 6061 at cryogenic temperatures.  Additionally, on the 4 K stage, 10-mils-thick high-purity (5N) annealed aluminum strips, purchased from ESPI Metals\footnote{http://www.espimetals.com/}, are used to create thermally-conductive parallel cooling paths along the lengths and circumferences of the 4 K radiation shields.  Despite its thin cross-sectional area---which allows it to easily mate to the curved geometry of the radiation shields---the material exhibits excellent thermal conductivity due to its high purity\cite{Woodcraft05}.  The strips are pressed and held with high pressure against the tubes through the use of large constant-tension hose clamps to maximize the contact force.  These hose clamps utilize heavy-duty springs along their tightening bolts, allowing them to maintain a strong tension and clamping force during thermal contraction on cool-down.  A thin layer of thermal grease was also applied between the radiation shields and high-purity aluminum strips before tightening the clamps.  This method avoids the use of any glues or epoxies, which can significantly degrade the effective thermal conductivity.

\subsubsection{Multi-layer insulation}

The heat load on the 50 K stage is dominated by radiative emission from the ambient-temperature vacuum shell, and multi-layer insulation blankets are used to decrease this power.  The blankets consist of 20 layers of aluminized-mylar sheets, thermally insulated from each other by a thin polyester material.  For durability, the blankets are taped along their edges using thin aluminized-mylar tape, and stapled with nylon clothing tags every 6 inches across their surface.

Although the use of multi-layer insulation is standard in most cryostats, initial cool-downs of the optics cryostat resulted in unexpectedly high temperatures and gradients along the 50 K radiation shields.  Through an iterative investigation, it was determined that the sources of unexpected heating were at the interface flanges between radiation shields (where there were small gaps in the insulation), and the blankets laid adjacent to each other.  Adding ``bridge-blankets" across these seams, which provided 3 inches of overlap between blankets, dramatically decreased the radiative load, causing the thermal gradient across the length of the 50 K optics tube to drop by $\sim$60 K and the total thermal load to drop by a factor of 2 (from $\sim$50 to 25 W).  This experience made it clear that the effectiveness of our multi-layer insulation was critically dependent on the handling of the overlap between blankets.

   \begin{figure} [ht]
   \begin{center}
   \begin{tabular}{c}
   \includegraphics[height=6.5cm]{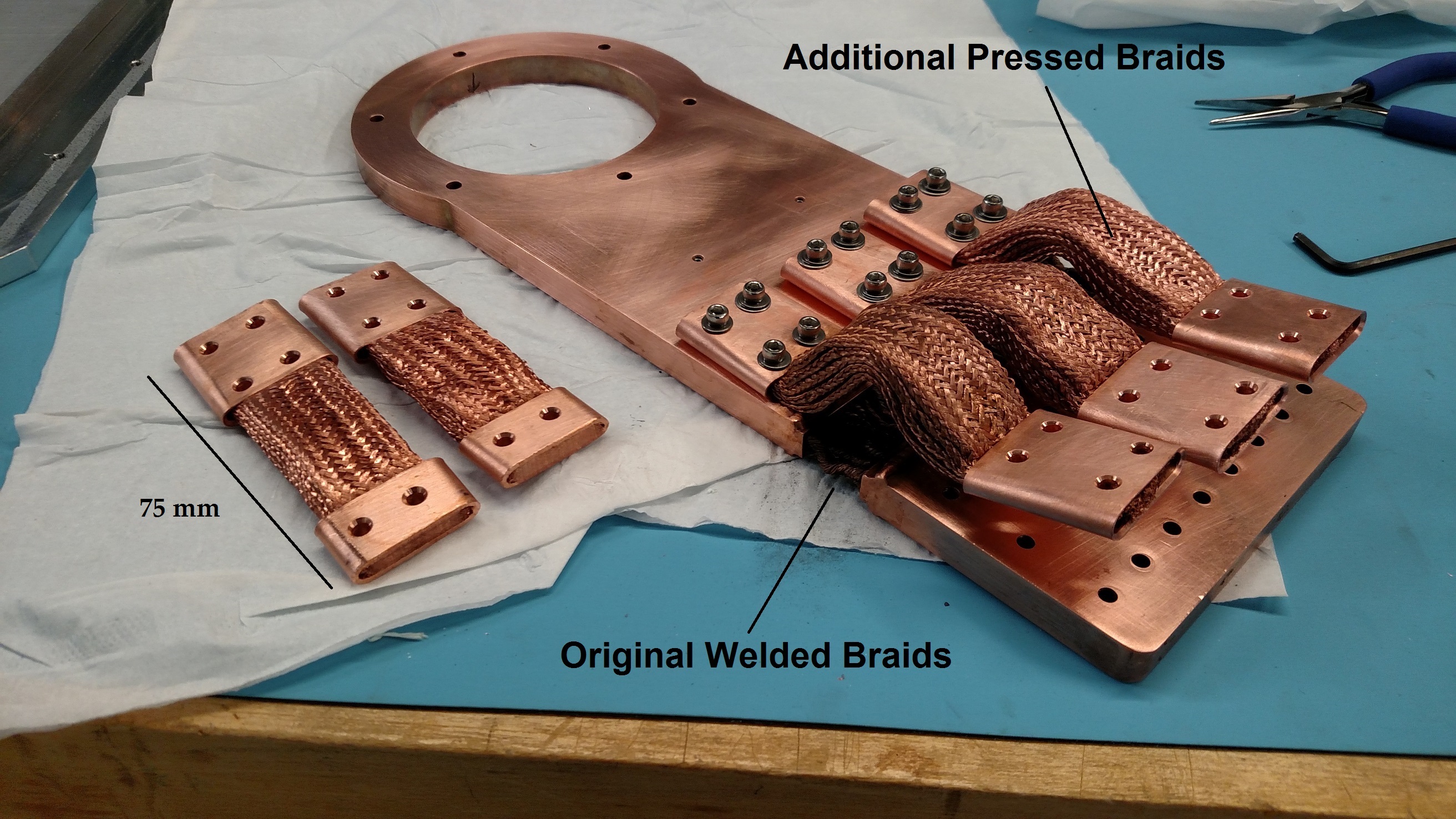}
   \end{tabular}
   \end{center}
   \caption[strap] 
   { \label{fig:strap} 
An example of the cryogenic copper thermal straps developed for SPT-3G.  Visible are both the original copper-welded braids, as well as the cold-pressed flat-braid straps developed and added in-situ to bolster the conductive throughput.}
   \end{figure}

\subsubsection{Cryogenic thermal straps}

Effectively coupling the cold-head of a cryogenic refrigerator to a dependent cold-stage is a common hurdle.  The movement of parts, due to thermal contractions and changes in pressure, demands that this thermal-mechanical coupling have physical flexibility.  Thermal straps in the SPT-SZ and SPTpol receivers relied on connecting flanges on the cold-head and cold-stage by welding an array of thin circular copper braids to both ends.  Welding copper and maintaining good conductivity is technically challenging due to the introduction of oxides and contaminants in the bond, and is a costly process due to the significant time, skill, and equipment it requires.  Furthermore, there is usually poor thermal contact between the braids and flanges, and the braids themselves can become rigid and fragile during the operation.

Similar straps were originally built for SPT-3G, but were later bolstered by additional straps made using a pressure-bond instead, as seen in Fig. 3.  Flat, annealed, OFHC copper braid purchased from Tranect Engineering Group\footnote{http://www.tranect.co.uk/} is cut to the desired length and layered before being inserted into short annealed OFHC copper tubes on both sides.  The inserted ends are then pressed with more than 10 tons of force to crimp the braid ends inside the flattened tubes.  The pressure produced during the operation is high enough to fuse the braid layers to each other and the inner surface of the tube with gas-tight bonds, ensuring that the mated surfaces do not oxidize and degrade over time.  The flattened ends can finally be machined with any desired hole geometry and fastened to each flange.

By avoiding the introduction of any heat to bond the materials, the braids do not experience rapid oxidation, and maintain their high-conductivity and flexible properties.  The strap ends effectively act as machinable monolithic copper blocks, which can be fastened to flange-surfaces with simplicity.  These straps are scalable in size/throughput and usable in a wide variety of geometries and temperatures.  Their implementation greatly improved the cryogenic performance of the SPT-3G receiver.  For example, Fig. 3 shows a strap used on the 50 K stage which normally experiences a 20 W load, and exhibited a 14.5 K gradient across the original welded braids.  The introduction of the three pictured flat braids decreased this gradient to only 5 K.  We hope to more systematically explore and document the performance of this thermal strap approach, but its qualitative success in the SPT-3G camera has already been met with interest by others in the CMB community due its simplicity and effectiveness\cite{Crumrine18}.

\begin{table}[ht]
\caption{Summary of predicted optical loading on detectors due to instrument and sky.  Calculated powers assume achieved temperatures during first-year observations.  Values in the total instrument row include contribution from unlisted filters as well.} 
\begin{center}
\begin{tabular}{|l|c|c|c|}
\hline
\textbf{}                 & \textbf{95 GHz (pW)} & \textbf{150 GHz (pW)} & \textbf{220 GHz (pW)} \\ \hline
Alumina Lenses            & 0.27                 & 0.55                  & 0.75                  \\ \hline
Lyot Stop                 & 0.40                 & 0.17                  & 0.02                  \\ \hline
Alumina Filter            & 0.24                 & 0.65                  & 1.09                  \\ \hline
HDPE Window               & 0.41                 & 1.26                  & 2.17                  \\ \hline
Telescope Mirrors         & 0.95                 & 1.84                  & 2.22                  \\ \hline
\textbf{Total Instrument} & \textbf{2.31}        & \textbf{4.50}         & \textbf{6.29}         \\ \hline
Atmosphere                & 2.67                 & 3.06                  & 3.65                  \\ \hline
CMB                       & 0.12                 & 0.12                  & 0.05                  \\ \hline
\textbf{Total Sky}        & \textbf{2.78}        & \textbf{3.18}         & \textbf{3.70}         \\ \hline
\end{tabular}
\end{center}
\end{table}

\subsection{Total Instrument Loading}

The mitigation of stray reflections onto hot, emissive surfaces and success in cooling optics components allows for relatively low instrument loading on the detector array.  During testing of the integrated detector and optics cryostat, the thermal gradient across the 2-m-long optics tube was measured to be 0.3 K.  The field lens, the alumina lens closest to the window, was measured to be 0.5 K hotter than the optics tube due to thermal gradients from infrared loading through the window and infrared filters, but regardless negligibly contributes to our in-band mm-wave loading.  All three alumina lenses operate at less than 5 K, and the Lyot stop is successfully held at less than 4.5 K.

Table 1 outlines the predicted optical loading which was used to set the saturation power of detectors\cite{Dutcher18}, including both instrument and sky sources.  As of the second year of observations, we have measured the ratio of measured to predicted optical power to be 1.07, 0.99, and 0.70 across the 95, 150, and 220 GHz detectors.  The fact that the measured optical power is not unexpectedly high builds confidence that the power emitted by the instrument on detectors is controlled and well understood.

\subsection{Optimizing Broadband Transmission}

Though minimizing instrument loading for SPT-3G was important, it was equally as important to design an optics system with high efficiency across the targeted millimeter-wave frequency bands (95, 150, and 220 GHz).  Low-loss microwave materials often have a high dielectric constant, meaning a significant amount of signal will reflect at every material boundary---leading to an exponential degradation in instrument efficiency with each additional element.  To minimize the cumulative loss in efficiency, anti-reflection (AR) techniques are frequently used on optics elements to improve their overall transmission.

However successfully developing a fieldable AR-technology was one of the biggest challenges in building the SPT-3G receiver.  Simple AR-coating methods typically only minimize reflections for a single frequency, and are not cryogenically viable due to thermal contraction differentials between the coatings, adhesives, and substrate.  AR-technologies with broader frequency-coverage are possible through the creation of surface micro-structures (which effectively result in frequency-dependent optical properties across a surface), or through the introduction of multiple coating layers with specifically chosen dielectric properties and thicknesses to create more complex interference dynamics between reflected waves.  The machinability of HDPE allows for triangular grooves to be cut into its surface relatively easily, a process which resulted in excellent and tunable broadband transmission across the surfaces of the vacuum window\cite{Raguin93}.  The alumina lenses posed a greater challenge due to the material's ceramic-composition and cryogenic requirements, and both thermal spray\cite{Jeong16} and PTFE-coating technologies\cite{Nadolski18} were developed and fielded during the first and second year of observations, respectively.  With both technologies, the measured optical efficiency to astronomical sources for the 220 GHz band was more than 50\% lower than predicted, whereas measured efficiencies roughly matched predictions for the 95 and 150 GHz bands.

   \begin{figure} [ht]
   \begin{center}
   \begin{tabular}{c}
   \includegraphics[height=5.75cm]{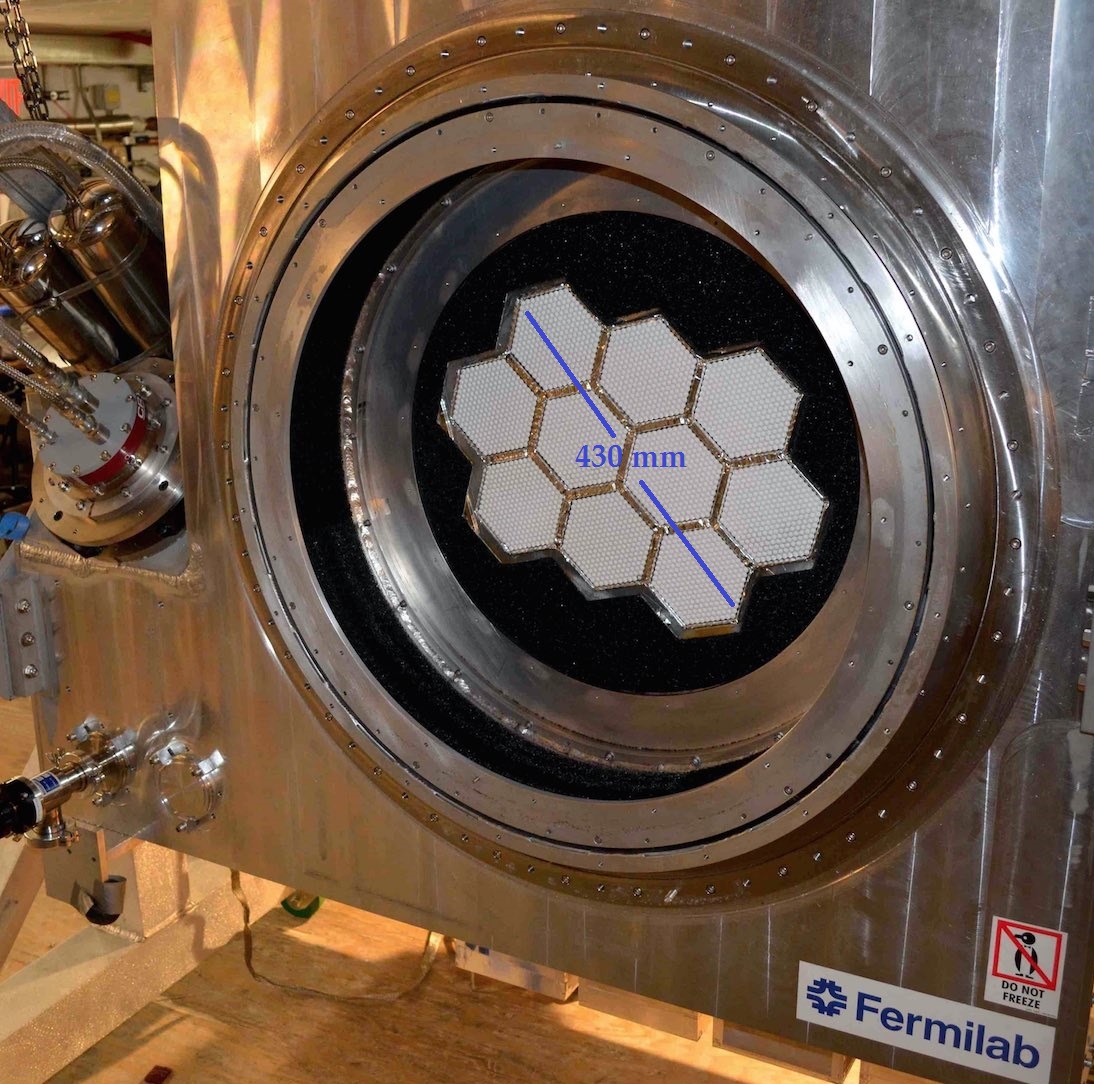}
   \includegraphics[height=5.75cm]{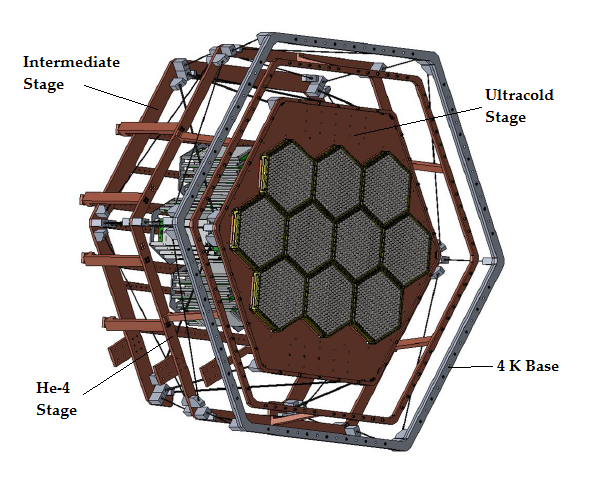}
   \end{tabular}
   \end{center}
   \caption[detectors] 
   { \label{fig:detectors} 
(Left) SPT-3G detector cryostat with detector array installed.  An absorptive aperture is matched to the non-circular footprint of the array.  (Right) A CAD image of SPT-3G sub-Kelvin architecture, including the detector array, cold-readout electronics, and thermal support stages.}
   \end{figure}

\section{DETECTOR CRYOSTAT}

The detector cryostat serves to cool the detector array and associated frequency-multiplexing readout system.  Details regarding fabrication of the SPT-3G transition-edge sensor bolometer array can be found in Ref. \citenum{Posada15}, and the most recent summary of properties and performance of the array can be found in Ref. \citenum{Dutcher18}.  Summaries of the frequency-multiplexing readout system can be found in Refs. \citenum{Bender16} and \citenum{Avva18}.

The detector cryostat includes a dedicated PTC which is positioned such that it remains close to vertical for optimal cooling efficiency when the receiver is in an observing orientation on the telescope.  Many of the cryogenic techniques which were described in Section 3 were utilized in the detector cryostat as well.  Though the temperatures of the detector cryostat components do not affect the instrument loading on detectors, SQUID amplifier performance improves with decreasing temperature.  The warmest SQUID amplifiers in the system, physically located the farthest away from the PTC, operate below 3.9 K, well within the necessary temperature range for low-noise performance.

SPT-3G detectors, along with their coupled inductor-capacitor resonators, must be cooled to sub-Kelvin temperatures to achieve photon-noise dominated sensitivities.  A closed-cycle \textsuperscript{4}He-\textsuperscript{3}He-\textsuperscript{3}He adsorption refrigerator purchased from Chase Research Cryogenics\footnote{http://www.chasecryogenics.com/} provides the necessary cooling power.  The refrigerator is a three-stage cooler capable of achieving temperatures below 250 mK at the ultracold-stage, and is designed to hold at base-temperature for 72 hours with 100, 10, and 2 $\mu$W on the three stages assuming negligible heat capacity.

Figure 4 shows the SPT-3G detector array and sub-Kelvin architecture.  This full assembly is installed through a large port in the rear of the cryostat.  The cold-stages are comprised of OFHC copper, and mechanically supported by carbon fiber reinforced polymer (CFRP) pultruded struts purchased from Goodwinds Composites.  This material exhibits an impressive strength to thermal conductivity ratio\cite{Runyan08} while being low-cost. Aluminized-mylar shielding (protecting the SQUID amplifiers from RF-contamination) and readout wiring are thermally coupled to the intermediate sub-Kelvin stages to decrease their heat loads on the ultracold stage.  The integrated sub-Kelvin assembly can hold at 270 mK for more than 24 hours during observations, with an observing efficiency of over 85\%, accounting for thermal cycling of the fridge and focal plane.

The detector array and cold-electronics for SPT-3G weigh over 50 pounds, making positioning them securely and stiffly while maintaining suitable thermal isolation between cold-stages challenging.  If the intrinsic vibrational modes of the assembly are matched to the vibrational modes of the telescope, excessive heating can occur on the sub-Kelvin stages due to microphonic vibrations via resonance with the telescope.  Though we see some evidence for microphonic sensitivity for certain sudden telescope movements, we have reduced these effects to a negligible level by adjusting the acceleration and velocity of the telescope during observations.

\section{Summary and Current Status}

We have described the design of the SPT-3G receiver.  In particular, we have described the cryogenic design and performance of the SPT-3G optics cryostat, which has demonstrated low thermal gradients, loading, and optics temperatures, through the use of cryogenic materials, multi-layer insulation, and new heat straps.  In addition, we describe the design and performance of the SPT-3G detector cryostat, which has also demonstrated sufficient cryogenic performance by achieving low SQUID temperature and high ($> 85\%$) observing efficiency.   The SPT-3G receiver achieved first-light in January of 2017, and began a 5-year survey of a 1500 deg$^2$ field in March of 2018. The successful implementation of optics, detector, and readout technologies has yielded a high sensitivity CMB receiver with significantly improved mapping speed over SPTpol.\cite{Bender18}

\acknowledgments 
 
The South Pole Telescope program is supported by the National Science Foundation (NSF)  through grant PLR-1248097.  Partial support is also provided by the NSF Physics Frontier Center grant PHY-1125897 to the Kavli Institute of Cosmological Physics at the University of Chicago, the Kavli Foundation, and the Gordon and Betty Moore Foundation through grant GBMF\#947 to the University of Chicago.  Work at Argonne National Lab is supported by UChicago Argonne LLC, Operator of Argonne National Laboratory (Argonne). Argonne, a U.S. Department of Energy Office of Science Laboratory, is operated under contract no. DE-AC02-06CH11357. We acknowledge R. Divan, L. Stan, C.S. Miller, and V. Kutepova for supporting our work in the Argonne Center for Nanoscale Materials.  Work at Fermi National Accelerator Laboratory, a DOE-OS, HEP User Facility managed by the Fermi Research Alliance, LLC, was supported under Contract No. DE-AC02-07CH11359.  NWH acknowledges support from NSF CAREER grant AST-0956135.  The McGill authors acknowledge funding from the Natural Sciences and Engineering Research Council of Canada, Canadian Institute for Advanced Research, and the Fonds de recherche du Qu\'ebec Nature et technologies.  Vieira acknowledges support from the Sloan Foundation.

We would like to thank A. Oriani for his thoughts and suggestions surrounding thermal strap development.

\bibliography{report} 
\bibliographystyle{spiebib} 

\end{document}